\documentclass[aps,preprint,floats,floatfix,tightenlines]{revtex4}

\usepackage{amssymb, amsmath, graphicx}

\begin{document}

\title{Relativistic Magnetic Monopole Flux Constraints from RICE}

\author{D. P. Hogan, D. Z. Besson, J. P. Ralston}
\affiliation{\small Department of Physics and Astronomy, University of Kansas, Lawrence, Kansas 66045, USA}
\author{I. Kravchenko}
\affiliation{\small Department of Physics and Astronomy, University of Nebraska, Lincoln, Nebraska 68588, USA}
\author{D. Seckel}
\affiliation{\small Department of Physics and Astronomy and Bartol Research Institute, University of Delaware, Newark, Delaware 19716, USA}

\begin{abstract}

We report an upper limit on the flux of relativistic monopoles based on the nonobservation of in-ice showers by the Radio Ice Cherenkov Experiment (RICE) at the South Pole. 
We obtain a 95\% C.L. limit of order $10^{-18}(\text{cm}^2\text{ s sr})^{-1}$ for intermediate-mass monopoles of $10^7\leq\gamma\leq10^{12}$ at the anticipated energy $E_\text{tot} =10^{16}$ GeV. This bound is over an order of magnitude stronger than all previously published experimental limits for this range of boost parameters $\gamma$ and exceeds 2 orders of magnitude improvement over most of the range.  We review the physics of radio detection, describe a Monte Carlo simulation including continuous and stochastic energy losses, and compare to previous experimental limits. 

\end{abstract}

\maketitle

\section{Introduction}

Magnetic monopoles are hypothetical particles carrying a net magnetic charge. Dirac\cite{Dirac31} calculated the quantum unit of magnetic charge while showing that existence of magnetic charge leads to charge quantization.  
 The relationship between the fundamental magnetic charge $g$ and the quantum of electric charge $e$ takes on the especially simple form 
$g=\frac{e}{2\alpha}$ in Gaussian units, where $\alpha$ is the fine structure constant.

 Magnetic monopoles occur in  grand unified theories (GUTs)\cite{thooft}.  Most GUTs predict a symmetry-breaking phase transition at an energy-scale $\sim10^{16}$ GeV\cite{Vilenkin94}.  Such a phase transition can lead to localized topological defects in the form of magnetic monopoles.  An order-of-magnitude estimate of one magnetic monopole per cubic Hubble radius gives a magnetic monopole number density at the GUT time of $n_M(t_{GUT})\approx 10^{82}$ m$^{-3}$, which would lead to magnetic monopoles dominating the Universe today\cite{Ryden03}, in conflict with observation.  Meanwhile, some inflationary models  
 predict dilution of GUT monopole density to hopelessly undetectable levels\cite{Guth81}.

No monopoles have been verifiably detected to date\cite{Eidelman04}.  Reports of magnetic monopole detections\cite{Price75, Cabrera82, Caplin86} have been challenged, often by the original authors themselves\cite{Price78, Huber90}.  Alternate explanations range from ionized nuclei\cite{Price78} to hardware malfunctions\cite{Caplin86} to new physics other than monopoles\cite{Price78}.

In this paper we present limits on relativistic monopole fluxes determined from five years of data collection by the RICE experiment\cite{Rice1}.  As noted by Wick {\it et al.}\cite{Wick03} the RICE array is particularly well-suited to ultrarelativistic monopole detection because of a combination of large effective volume and favorable scaling with energy.  Our limits for fluxes over the range of monopole Lorentz boost parameters $ 10^8 \lesssim \gamma \lesssim 10^{12}$ are about 100 times more restrictive than the tightest published limits from AMANDA\cite{Wissing07}, Baikal\cite{Aynutdinov05}, and MACRO\cite{Ambrosio02}, and more than 1000 times more restrictive than the original Parker bound. 

This paper first reviews monopole properties and previous limits in Sec. \ref{sec:monoreview}.  We then provide an overview of RICE in Sec. \ref{RICEsection}.  Sections \ref{sec:eloss}, \ref{sec:MC}, and \ref{sec:response} discuss monopole energy loss, our Monte Carlo simulation thereof, and RICE's response.  Results are summarized in Sec. \ref{sec:results}.

\section{Light relativistic monopoles}
\label{sec:monoreview}

\subsection{Formation}

Until monopoles are detected their masses should be considered unknown.  Although GUT-scale monopoles are commonly believed to be extremely heavy ($10^{17}$ GeV) and undetectably rare as a result of inflation, there are other mechanisms resulting in production of much lighter monopoles after inflation.  Kephart and Shafi\cite{Kephart01} proposed that later symmetry-breaking events at lower energies could occur, resulting in monopoles with a magnetic charge of a few Dirac units and masses in the range $10^{7}-10^{13}$ GeV.  Other candidate GUTs lower the GUT energy scale, allowing for $10^8$ GeV monopole mass ranges, and conformal field theories can be developed allowing for monopole masses on the order of $10^5$ GeV\cite{Wick03}.  Because magnetic charge is conserved, magnetic monopoles formed in the early universe should persist through the current epoch\cite{Giacomelli01}.  In fact, these lighter monopoles have been suggested as possible ultra-high energy cosmic rays  beyond the GZK\cite{GZK} cutoff\cite{Wick03}.

Such considerations motivate the search for so-called intermediate-mass monopoles (IMM's), that is, monopoles with mass significantly less than the conventional GUT energy. Our study concentrates on this mass range because such monopoles are expected to be ultrarelativistic and readily detectable by the RICE array.  

\subsection{Experimental Limits To Date}

In 1970, Parker pointed out that the abundance of magnetic monopoles is constrained by the requirement that magnetic monopole currents be insufficient to deplete the galactic magnetic field\cite{Parker70}.  Only in the past decade have astrophysical experiments been able to improve upon the original Parker bound of flux $\sim 10^{-15}(\text{cm}^2\text{ s sr})^{-1}$\cite{Ambrosio02}.  The first observational astrophysics experiment to obtain limits stronger than the Parker bound was MACRO (the Monopole Astrophysics and Cosmic Ray Observatory) at Gran Sasso, Italy. The MACRO bound for monopole velocities $v =\beta c>0.99c$ is a flux upper limit of $1.5\times10^{-16}(\text{cm}^2\text{ s sr})^{-1}$.  Upper bounds of this order of magnitude were also reported for $4\times10^{-5}<\beta<0.99$\cite{Ambrosio02}.

Since MACRO's termination in 2000, ``neutrino telescopes'' have conducted searches for relativistic magnetic monopoles.  Both AMANDA (the Antarctic Muon and Neutrino Detector) 
 and the Baikal Neutrino Telescope have reported flux limits stronger than MACRO's for $\beta\geq0.8$.  Baikal has set an upper bound on monopole flux of $5\times10^{-17}(\text{cm}^2\text{ s sr})^{-1}$ for $\beta\approx1$ and $10^7\leq M\leq10^{14}$\cite{Aynutdinov05, Baikalnote}.  AMANDA's most recent limit for monopole masses $10^8\leq M\leq10^{11}$ is $2.8\times10^{-17}(\text{cm}^2\text{ s sr})^{-1}$ at $\beta\approx1$\cite{Wissing07}.  Preliminary results from IceCube's 9-string configuration indicate that new lower bounds [somewhat below $10^{-17}(\text{cm}^2\text{ s sr})^{-1}$ at $\beta\approx1$] may be forthcoming\cite{ninestring}.

The SLIM (Search for Light Magnetic Monopoles) experiment at the Chacaltaya High Altitude Laborary in the mountains of Boliva offers complementary sensitivity to IMM's.  This nuclear track detector experiment is designed especially to search for light monopoles (mass $10^5$ GeV to $10^{12}$ GeV) over a wide range of velocities (including $\beta\geq4\times10^{-5}$ for 1-Dirac-charge monopoles).  SLIM's latest monopole flux limit at $\beta\approx1$ is $6.5\times10^{-16}(\text{cm}^2\text{ s sr})^{-1}$ for Earth-crossing monopoles, or $1.3\times10^{-15}(\text{cm}^2\text{ s sr})^{-1}$ if upgoing monopoles are blocked\cite{Balestra08}.

Stronger flux limits based on astrophysical considerations [such as an ``extended Parker bound'' of less than $3\times10^{-22}(\text{cm}^2\text{ s sr})^{-1}$ for IMM's, based on a more realistic model of galactic magnetic fields\cite{Adams93}] have also been proposed.

\subsection{Relativistic IMM's}
Because of their moderate mass, IMM's may acquire highly relativistic velocities.  Wick \emph{et al.}\cite{Wick03} use a model of monopole traversal of intergalactic magnetic fields (similar to the model underlying the Parker bound) to estimate that IMM's created in the early universe would now have typical kinetic energies on the order of $10^{16}$ GeV, with a comparable spread in energy.  PeV-mass monopoles would therefore reach boost factors $\gamma\approx 10^{10}$.
The fact that IMM's acquire such ``ultrarelativistic'' $\gamma$ values provides a mechanism for their detection.  Any particle travelling through a medium loses energy, but ultrarelativistic charged particles do so dramatically, initiating showers in the surrounding medium\cite{Jackson62}.  It is through detection of such showers that RICE is sensitive to magnetic monopoles.

\section{RICE} \label{RICEsection}

RICE (the Radio Ice Cherenkov Experiment) consists of 16 data-taking antennas buried in the Antarctic ice at the Martin A. Pomerantz Observatory (MAPO) about 1 km from the geographic South Pole.  The antennas, which are roughly contained within a cube of ice $\sim$200 m on a side with its center $\sim150$ m below the surface, have peak sensitivity in the 200--500 MHz regime.  Collisions of highly energetic neutrinos ($E>10^{8}$ GeV) with in-ice atomic nuclei result in  electromagnetic and/or hadronic cascades, depending on neutrino flavor\cite{Rice1}. The superluminal velocity of the charged particles comprising the cascade  creates coherent broadband Cherenkov radiation in 
the radio frequency domain.  The broad frequency spectrum leads to a sharp detectable pulse of Cherenkov radiation\cite{Razzaque02,Rice1, Rice2} in the time domain.  The pulse  propagates outward from the shower axis in a cone with opening angle $\theta_c$ given by \cite{Jackson62}
\begin{equation} \label{cangle}
\theta_c = \arccos\left(\frac{1}{\beta n}\right).
\end{equation}
Here $n$ is the index of refraction of the medium evaluated at the radiation frequency.   For highly relativistic particles ($\beta\approx 1$) in ice ($n \approx 1.78$), the Cherenkov angle is about 0.97 rad.

\subsection{RICE hardware and data taking}

We now briefly summarize the RICE operation, referring the reader to more detailed descriptions found elsewhere\cite{Rice1}.  

When the RICE detector is live, a trigger occurs if four or more antennas register high-amplitude voltages within a time coincidence of 1.25 microseconds.  Triggers initiate an 8.192 microsecond waveform capture, sampled at a rate of $1\times 10^{9}$ samples per second, for all under-ice antennas. Wave forms are approximately centered on the time that the trigger latched the data acquisition system. Software surface vetoes (maximum rate $\sim$80 Hz) and hardware surface vetoes (maximum rate $\sim$200 kHz) suppress anthropogenic backgrounds originating above the array on the surface. Offline, the signal profiles are fed through a reconstruction algorithm, described in\cite{Rice3}, which is designed to reject spurious triggers while preserving a large fraction of high-energy cascade signals.  During data collection from 1999 to 2005, RICE was operated for a livetime of $74.1\times10^6$s, resulting in $1.035\times10^6$ triggers.  All triggered events were ruled out as viable neutrino candidate events during offline analysis, from which an upper bound on the diffuse high-energy neutrino flux was derived\cite{Rice3}.  Because relativistic monopoles also generate in-ice cascades, this same data set can be used to derive an upper bound on the flux of relativistic monopoles.  To produce a bound it is necessary to quantify the sensitivity of RICE to monopoles.  A Monte Carlo simulation code (``monoMC'') was therefore created to evaluate RICE's monopole detection efficiency, as detailed below.

\section{Monopole energy loss in matter}
\label{sec:eloss}

Our model of monopole energy loss is based on the muon/tau energy loss model of Dutta \emph{et al.}\cite{Dutta01}.  In this model, energy loss by a muon traversing a medium is expressed as
\begin{equation} \label{losssummary}
-\frac{dE}{dx} = \alpha + \beta E.
\end{equation}
The $\alpha$ term is the energy loss per distance (units: g/cm$^2$) due to ionization of the medium.  The $\beta$ term\cite{notebeta} is the sum of three terms reflecting bremsstrahlung, pair production, and photonuclear effect energy losses.  Each energy loss mechanism is calculated separately.  Defining $y$ to be the fraction of its energy lost by the particle in a single interaction with the medium, each of the three terms $\beta_i$ is found by convolving $y$ with the partial interaction cross section with respect to $y$ \eqref{inty}:
\begin{equation} \label{inty}
\beta_i(E) = \frac{N}{A} \int_{y_{\text{min}_i}}^{y_{\text{max}_i}} y \frac{d \sigma_i(y,E)}{dy} dy
\end{equation}
Here, $N$ is Avogadro's number and $A$ is the average atomic mass number of the medium.  The full formulae for $\alpha$ and the $y_{\text{min}_i}$, $y_{\text{max}_i}$, and ${d \sigma_i}/{dy}$ needed to calculate each $\beta_i$ are given in\cite{Dutta01}.  In general, the expressions are functions of particle mass and energy and of various properties of the medium, although $\alpha$ and the individual $\beta_i$'s are only weakly energy dependent.

\subsection{Discrete Loss Processes}

Although energy loss due to ionization can be treated as smooth and continuous with little loss of accuracy, we explicitly model the stochastic fluctuation in pair production and photonuclear energy losses.  Combining Eq. \eqref{losssummary} with \eqref{inty} and replacing the integral in \eqref{inty} with the corresponding Riemann sum gives the result \eqref{Rsum}, where $\Delta E_i$ is the energy loss via process $i$ (brem., pair, or photonuclear) over a small distance $\Delta x$:
\begin{equation} \label{Rsum}
\Delta E_i \approx \sum_{ \substack{j \\ y_j=y_{\text{min}_i}} }^{y_j=y_{\text{max}_i}}\left(\frac{N}{A}\right)(y_j E)\left( \Delta x \frac{d\sigma_i}{dy_j}\right)\Delta y
\end{equation}
Recasting the energy loss equation this way effectively sorts the total energy loss into an arbitrary number of bins, each of which spans a length $\Delta y$ of the possible $y$ values.  Since $y_j$ is the fractional energy loss in a single interaction within bin $j$ and $E$ is the total energy of the particle, $(y_j E)$ is the energy loss for a single interaction in the $j^{\text{th}}$ bin.  Each term of the Riemann sum represents an energy loss, so if $(y_j E)$ is the energy lost in a single interaction, the remaining multiplicative terms in the summand give the expectation number of interactions in the $j$th bin $\langle n_{ij} \rangle$:
\begin{equation} \label{expectation}
\langle n_{ij} \rangle = \frac{N}{A}\Delta x \frac{d\sigma_i}{dy_j}\Delta y
\end{equation}
Therefore, accurately modeling the stochastic variation in bremsstrahlung, pair production, and the photonuclear effect is equivalent to replacing $\langle n_{ij} \rangle$ by a random number drawn from a Poisson distribution of expectation value $\langle n_{ij} \rangle$ when evaluating the energy loss expressions \eqref{Rsum}.

\subsection{Generalization to Monopoles}
\label{sec:model}

Only a few changes are needed to convert  the stochastic model of muon energy loss to a model of magnetic monopole energy loss.  First, the muon mass must be replaced by the magnetic monopole mass.  Because bremsstrahlung falls off by inverse powers  of particle mass, the bremsstrahlung energy loss contribution is negligible for even light magnetic monopoles and will be subsequently disregarded\cite{Wick03}.  It should be noted that at large masses ($\gtrapprox$1 TeV), $\beta_\text{pair production}$ can become difficult to calculate numerically due to rounding error.  However, pair production energy loss approaches an asymptotic limit with increasing particle mass and varies with mass by only a few percent for masses above $\sim 100$ MeV.

Next, due to Dirac's quantization condition a magnetic monopole of 1 Dirac charge will lose energy equivalent to an electric charge of $1/(2\alpha)$ times the proton charge\cite{Jackson62}.  Accounting for this large effective charge only
requires multiplying the expectation number of interactions by $1/(2\alpha)^2 \approx 4700$.

This procedure for modelling magnetic monopole energy loss in matter has assumed the particle to be a simple Dirac monopole, that is, a point source of magnetic charge with no further internal structure.  ``Actual'' magnetic monopoles may contain internal color fields and lose additional energy through hadronic interactions beyond the photonuclear effect\cite{Rubakov}. However calculations of such processes are highly model-dependent\cite{Wick03} and not further considered in this analysis.

\begin{figure}\centering
\includegraphics[width=0.5\textwidth]{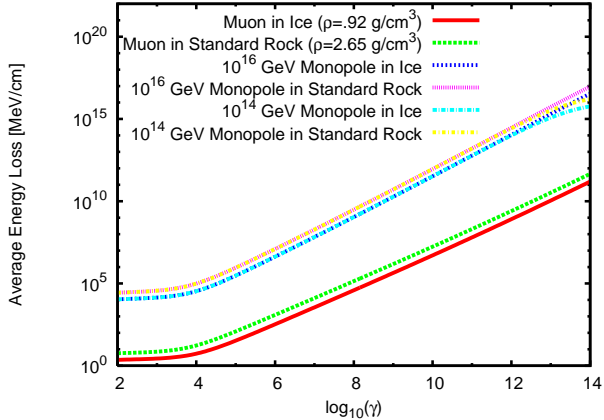}
\caption{(Color online) Muon energy loss in ice and standard rock compared to monopole energy loss in ice and standard rock, as a function of boost parameter $\gamma$.  Monopole results are shown for total energy $10^{16}$ GeV (the energy assumed in this analysis) and, for comparison, $10^{14}$ GeV.  The figure shows that, over the kinematic range of interest, monopole energy loss depends strongly on $\gamma$, but for a given $\gamma$ it is only weakly mass dependent.}
\label{energyloss}
\end{figure}

\begin{figure}
\centering
\includegraphics[width=0.5\textwidth]{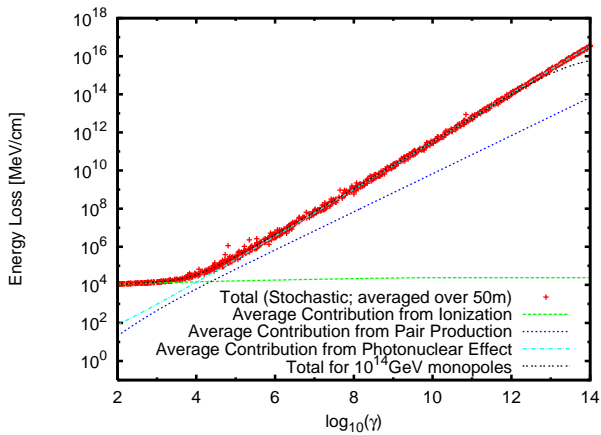}
\caption{(Color online) Total energy loss versus $\gamma$ for $10^{16}$ GeV monopoles, showing stochastic variation over 50m intervals.  Lines show average contributions from different processes.  Also shown, again for comparison, is the average total energy loss for $10^{14}$ GeV monopoles.  In the latter case, stochastic variation and energy loss contributions from the various processes are almost identical to the former case.}
\label{monobyparts}
\end{figure}

Figure \ref{energyloss} shows the average monopole energy loss in ice and ``standard rock'' (A=22, $\rho=2.65 \text{~g/cm}^3$\cite{Dutta01}), along with the same results for muons.  While the muon mass is fixed, the monopole rest mass is constrained to vary inversely with gamma such that total energy is fixed at a reference energy of $10^{16}$ GeV.   Figure \ref{monobyparts} shows various contributions to the energy loss of a  $10^{16}$ GeV monopole. Much of the difference between muons and monopole energy losses is due to the monopole's large effective charge.   The curves indicate average energy loss due to the three principal mechanisms, while the points show actual stochastic energy loss (as averaged over a 50 m interval).  The photonuclear effect is the dominant energy loss mechanism at $\gamma>10^4$, while ionization energy losses dominate below this value\cite{noteerror}.  Because the photonuclear mechanism results in hadronic showers generated by nuclear recoils, we may ignore LPM\cite{LPM} effects.

There are substantial uncertainties in extrapolations of photonuclear losses to ultrahigh energies.  The primary unknown is the hadronic contribution of real photon-nucleon scattering. Consider an extrapolation based on the Froissart bound setting in at about 50 GeV\cite{Froissart}.  The energy dependence of the photon-nucleon cross section $\sigma_{\gamma N}$ of such a model is \begin{eqnarray} \sigma_{\gamma
N}(E_{\gamma}) = 114.3+1.67 \,{\rm ln}^2 (0.0213 E/{\rm GeV})\, \mu{\rm b}
\label{bb-photonuclear} 
\end{eqnarray} 

This 1981 model of Bezrukov and Bugaev\cite{Bezrukov81} predates the discovery at HERA of parton distributions at small$-x$ that causes cross sections to grow like fractional powers.  For comparison, the 1998-2001 post-HERA photonuclear cross section of Donnachie and Landshoff \cite{DL} is within 10\% of Eqn. \eqref{bb-photonuclear} at $E_{\gamma}=10^{6}$ GeV, while being about 16 times larger at $10^{11}$ GeV.  As shown in Fig. 3 of Ref. \cite{Dutta01}, the more complicated cross sections developed by extrapolating structure functions track the simple expression of Eq. \ref{bb-photonuclear} very well.  
Using small and slow-growing cross section models such as these is in some sense conservative.  It results in dim showers that are less likely to trigger.  However, small cross sections also tend to develop fewer showers failing the time-over-threshold cut discussed below, and fewer showers are stopped by the Earth.  No one cross section model is in all cases the ``most conservative.''

\section{Monopole Monte Carlo}
\label{sec:MC}

The monopole energy loss model described above has been incorporated into a Monte Carlo simulation of a magnetic monopole travelling near RICE.  The Monte Carlo simulation randomly generates a monopole trajectory and energy loss, then determines the voltage response of the RICE antennas.  Each voltage profile is subsequently processed using the same reconstruction software as was used in the 2006 RICE data analysis\cite{Rice3}.

With the Monte Carlo simulation, we generate monopoles at the reference energy ($10^{16}$ GeV) with seven $\gamma$ values logarithmically spaced from $10^6$ to $10^{12}$.  The maximum value of $\gamma=10^{12}$ corresponds to a 10 TeV monopole mass.  Smaller masses are probably unrealistic theoretically.  The lower $\gamma$ bound is somewhat ad hoc, chosen to bracket the kinematic regime to which RICE is most sensitive.

For each simulated monopole, the Monte Carlo first generates a random monopole trajectory with an impact parameter (distance of closest approach to the array) less than a $\gamma$-dependent maximum impact parameter $r_m$\cite{noteorigin}.  The $r_m$ value used for a given simulation series must be large enough to include virtually all detectable monopoles without being so large that zero or a negligible number of the simulated monopoles are detected, in which case no flux bound can be calculated (or the flux bound will be unnecessarily high after accounting for statistical error).  The selected $r_m$ values were based on preliminary simulations and are shown in Table \ref{tableflux}.

\subsection{Passage through Earth}

Unlike ultra-high energy neutrinos, not all upcoming monopoles range out before reaching the detector, although their energy loss in-transit can be substantial.  The terrestrial density integrated over the length of the monopole's path [g/cm$^2$], as a function of approach angle, is taken from the Preliminary Reference Earth Model\cite{Dziewonski81}.  The monopole energy loss in transiting this material is calculated over 50 increments of equal column thickness.  The calculation assumes $A=22$ for material within the Earth; however, the photonuclear effect, which dominates energy loss, is independent of $A$ aside from a nuclear shadowing factor affecting the result at the level of a few percent.  If at any time the monopole's $\gamma$ falls below 2, the monopole is considered to have effectively stopped.  In practice much larger $\gamma$ values are needed for RICE to trigger.

\begin{figure}
\centering
\includegraphics[width=0.5\textwidth]{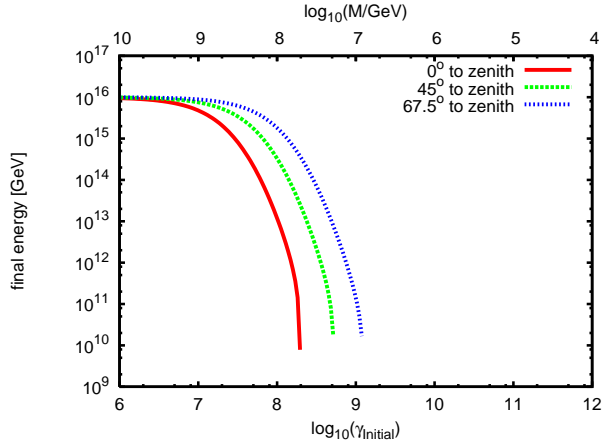}
\caption{(Color online) Final energy of magnetic monopoles with initial energy $10^{16}$ GeV after crossing the Earth.  The angle labeling each curve is the opening angle between the monopole's velocity vector and the zenith at the monopole's point of exit.}
\label{earth_gamma}
\end{figure}

Figure \ref{earth_gamma} shows the energy remaining after crossing the Earth for monopoles initially having energy $10^{16}$ GeV with a range of incidence angles.  Energy loss increases with $\gamma$, so monopoles with $\gamma\lessapprox 10^7$ lose a negligible fraction of their incident energy.  At higher $\gamma$, energy loss can be significant.    Beyond $\gamma\gtrapprox 10^{10}$, the Earth is opaque to monopoles.

\subsection{Propagation Through Ice} \label{Propagation}

After calculating the energy lost by a monopole en route to the ice target, the monopole's interaction with the ice itself is simulated.  So long as the monopole is ``out of range'' of the RICE antennas, it is propagated along in 10 m steps, with its $\gamma$ value decremented in each step to account for energy lost in travelling that 10 m path.  A monopole is considered to be ``in range'' as long as at least one of the RICE antennas is within 0.33 rad of the Cherenkov cone anywhere along a 900 m segment extending from 550 m ahead of the monopole to 350 m behind it.  (We choose an extended path length to accurately model the time-over-threshold cut in RICE's offline analysis, described later.)  The angle 0.33 rad is 1.70 half widths of the Cherenkov radiation's angular distribution at 0.2 GHz\cite{Alvarez00,Razzaque02}, which is the lower edge of RICE's frequency sensitivity.  At higher frequencies the Cherenkov radiation's angular distribution is more tightly confined.  The path cutoff distances of 550 m and 350 m were chosen to guarantee that all signal arriving at an antenna in the first 1.2 $\mu$s following the Cherenkov peak will be considered ``in range''\cite{noterange}.

Once a monopole comes into range, the size of the simulation steps is reduced from 10 m to 0.4 m.  As a result of this small step size, all the energy lost within the interval can be treated as originating at a single ``subshower'' with a pointlike source, while introducing signal arrival timing errors no greater than 0.4 ns.  By comparison, the actual experiment's digitizer samples every nanosecond. 

The complete voltage profile $V(t)$ at each antenna is determined by coherently summing the voltage contributions of various subshowers in each time bin.  Since the photonuclear effect dominates energy loss, the initial energy of each subshower is taken to be equal to the monopole's total energy loss within the corresponding distance interval.  The signal phase will vary slowly with viewing angle\cite{buniy}; however, in most cases the dominant emission arises from a coherent region along the track centered around the Cherenkov point.  This region is typically of length $\sqrt{\lambda R} \sim 10-50$~m and subtends a small viewing angle.  In rare cases, single large subshowers at modest distances dominate the radiation, but in this case the absolute phase of a single subshower is irrelevant.  Accordingly, we ignore all signal phases other than from travel time in our analysis.

\section{RICE response}
\label{sec:response}

\subsection{Signal Analysis}

RadioMC, a Monte Carlo simulation of the radiofrequency signals caused by cascades (discussed in Ref.\cite{Rice3}) has been previously developed in the context of RICE's high-energy neutrino flux studies.  
Because of the necessity of modeling a large number of subshowers for each monopole, we use a streamlined version of RadioMC that does not take into account the depth-dependence of the index of refraction in the firn, defined as the upper 175 m of ice.  This introduces two errors for which we correct as follows:  First, in order to account for the presence of a ``shadow zone''\cite{Rice3}, voltage contributions are suppressed when a monopole is more distant than a depth-dependent limiting horizon.  Although this horizon increases with both source depth and antenna depth, it blocks all but two of the antennas from seeing even the deepest events at distances greater than 20km.  
Second, the streamlined code overestimates signal arrival time by assuming that firn ice has the same index of refraction as deep ice ($n=1.78$), whereas $n$ actually falls off to $1.37$ at the surface.  We estimate the timing error from each subshower's signal using a parameterization of index of refraction vs depth, and time-shift each subshower's voltage contribution accordingly.  This effect translates the Cherenkov point along the path, as a first-order correction for refractive effects.

\begin{figure}
\centering
\includegraphics[width=0.5\textwidth]{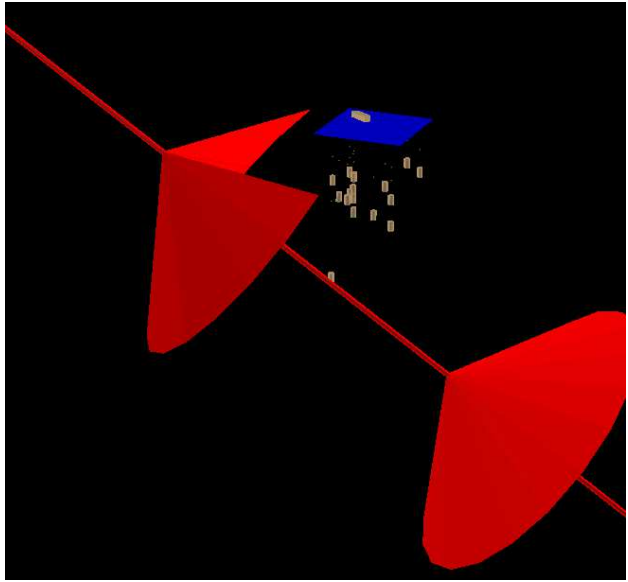}
\caption{(Color online) Path geometry of a nearby, but otherwise typical, downgoing monopole ($\gamma=10^7$).  See explanation in text.}
\label{3Dpicture}
\end{figure}

\begin{figure}
\centering
\includegraphics[width=0.5\textwidth]{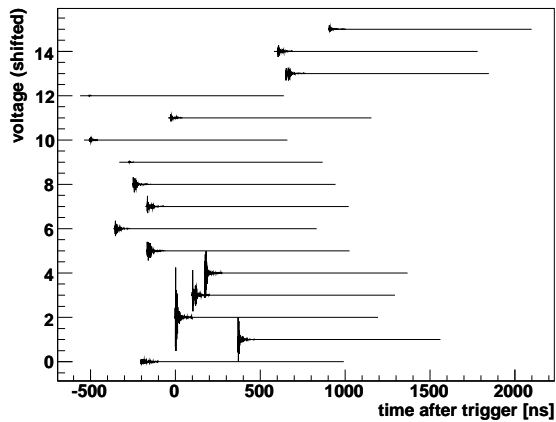}
\caption{Simulated voltage vs time (as measured at the data acquisition system) in each RICE antenna channel as caused by the monopole of Fig. \ref{3Dpicture}.  The voltage graphs for the different antennas have been shifted vertically for clarity.}
\label{voltgraph}
\end{figure}

From the voltage profiles generated by RadioMC, we determine which monopoles would cause the array to trigger.  RICE triggers when four or more antennas exceed a voltage threshold of roughly 200--400 mV (after amplification, and depending on the background levels at a given time) within 1.25 $\mu$s.  As the precise threshold varied over the life of the experiment, the threshold used in any given Monte Carlo simulation is randomly selected from the historical threshold distribution.  A typical example of a ``trigger'' is shown in Figs. \ref{3Dpicture} and \ref{voltgraph}.  In Fig. \ref{3Dpicture}, the long, diagonal line indicates the downgoing monopole's path, and the cones show the Cherenkov radiation emitted at the boundaries of the ``in range'' path segment.  The missing section of the first Cherenkov cone (at upper-left) indicates where the cone intersects the ice surface.  The square corresponds to the surface area mapped in Fig. 3 of\cite{Rice1}, with the RICE antennas (enlarged for clarity) arranged in the ice below.  For scale, the MAPO building shown on the surface is $\sim$50 m long.  Figure \ref{voltgraph} shows the voltage signal generated by the same monopole.

As a final analysis step, following the procedure of\cite{Rice3}, each calculated monopole voltage profile is embedded in an unbiased event (a random recording from the actual RICE antennas, including thermal noise, etc.).  The combined signal is fed through the same event reconstruction algorithm used in the 2006 neutrino analysis, to determine which monopoles would survive the offline event reconstruction analysis.

\subsection{Flux Upper Bounds}

Supposing that a number $n$  from a sample of $N$ monopoles trigger the array and survive reconstruction in a simulated sample with maximum impact parameter $r_m$, an upper bound on monopole flux is derived as follows.  Although the literal detection efficiency is $\epsilon_t=\frac{n}{N}$, we take the efficiency to be the lower bound of a 90\% Agresti-Coull confidence interval\cite{Brown01} about $\epsilon_t$.  Then, $\epsilon_s$ has only a 5\% chance of being greater than the efficiency that would be obtained in an infinite number of trials.  The cross section for monopole detection is
\begin{equation} \label{crosssection}
\sigma = \pi r_m^2 \epsilon_s f_b.
\end{equation}
Here $f_b$ is a factor which reflects the estimated decrease in detection due to the birefringence effect, which is not directly modeled in the Monte Carlo\cite{Besson07}.  The 0.12\% birefringence implied by measurements near Taylor Dome correspond to a sensitivity reduction of approximately 14\% that is roughly independent of energy; recent measurements at the South Pole indicate no observable birefringence ($<$0.01\%) to a depth of $\sim$1.1 km\cite{Besson08}.  We conservatively take $f_b$=0.86 for the purposes of calculating our upper limit.

Assuming Poisson statistics, the 95\% upper bound on monopole flux is
\begin{equation} \label{ubound}
F_b = \frac{2.995}{4\pi L \sigma}.
\end{equation}
The factor $4\pi$ converts the flux from units of $(\text{cm}^2\text{ s})^{-1}$ to $(\text{cm}^2\text{ s sr})^{-1}$, assuming an isotropic distribution of monopoles; $L$ is the livetime of the experiment.  Although the 2006 RICE neutrino analysis incorporated data from 1999 through 2005, the 1999 and 2000 data are not considered for this monopole analysis due to differences in the detector configuration between those earlier data and subsequent ($>$2000) datasets.  Thus we use a livetime of $58.3\times 10^6\text{s}$, corresponding to the 2001-2005 RICE operations, for calculation of our sensitivity.

\section{Results}
\label{sec:results}

\begin{figure}
\centering
\includegraphics[width=0.5\textwidth]{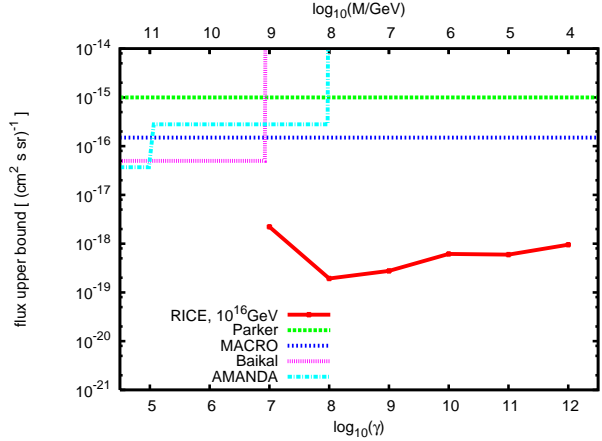}
\caption{(Color online) Upper bound on magnetic monopole flux from RICE's null observation in 2001-2005 versus monopole mass and $\gamma$.  Also shown are the Parker bound and results from MACRO\cite{Ambrosio02}, AMANDA\cite{Wissing07}, and Baikal\cite{Aynutdinov05}.}
\label{figflux}
\end{figure}

Figure \ref{figflux} shows the upper bound on magnetic monopole flux as a function of $\gamma$ for monopoles of initial energy $10^{16}$ GeV.  
These limits are based on a few to a few dozen simulated detections per $\gamma$ out of a generated sample of ten thousand monopoles per $\gamma$, as tabulated in Table \ref{tableflux}
\cite{notebound}.
Our limits have been degraded by the statistical uncertainty in our Monte Carlo-estimated efficiency for each $\gamma$ bin.

\begin{table*}
\begin{tabular}{|c|c|c|c|c|c|c|}
\hline
$\text{log}_{10}(\gamma)$ & $\text{log}_{10}(\frac{\text{mass}}{\text{GeV}})$ & \#sim. & $r_m$ / km & \#detec. & $\frac{\sigma}{\text{km}^2}$  & $\frac{\text{flux bound}}{10^{-19}(\text{cm}^2\text{ s sr})^{-1}}$ \\
\hline
~6 & 10 & $1\times10^4$ & ~3 & ~0   & $<$0.0073  & $>$560~~~~ \\
~7 & ~9 & $1\times10^4$ & ~3 & 90   & 0.18  & 22~ \\
~8 & ~8 & $1\times10^4$ & ~5 & 342~ & 2.1~ & 1.9 \\
~9 & ~7 & $1\times10^4$ & 10 & 67   & 1.5~ & 2.8 \\
10 & ~6 & $1\times10^4$ & 12 & 24   & 0.66  & 6.2 \\
11 & ~5 & $1\times10^4$ & 14 & 19   & 0.68  & 6.0 \\
12 & ~4 & $1\times10^4$ & 14 & 13   & 0.43  & 9.5 \\
\hline
\end{tabular}
\caption{Final Monte Carlo results as a function of monopole $\gamma$ and mass (columns 1 and 2).  Subsequent columns show Monte Carlo parameters (number of simulations, maximum impact parameter $r_m$, number of monopoles passing software reconstruction, and the resulting cross section for monopole detection by RICE).  The resulting bound on monopole flux is tabulated in column 7 and plotted in Fig. \ref{figflux}.}
\label{tableflux}
\end{table*}

We note that across a range spanning 4 orders of magnitude in $\gamma$, the flux of $10^{16}$ GeV, single-Dirac-charge monopoles incident on the Earth is less than 1$\times10^{-18} (\text{cm}^2\text{ s sr})^{-1}$.  
Below $\gamma\approx10^8$, the utility of RICE for IMM detection rapidly deteriorates.

For a given initial energy (as assumed here), several factors contribute to the $\gamma$ dependence of the upper bound on monopole flux.  
Monopole energy loss, and hence signal strength, rises with $\gamma$.
Below $\gamma\approx10^7$, few monopoles are bright enough to cause the array to trigger.  Sensitivity peaks at $\gamma\approx10^8$ and falls off again at higher $\gamma$.  The blocking of upgoing monopoles for $\gamma\geq 10^{10}$ contributes to this high-$\gamma$ sensitivity falloff.  
However, a more significant effect at $\gamma\gtrapprox10^9$ is decreasing efficiency as more monopoles fail the ``time-over-threshold'' cut.  This cut is a requirement in the offline analysis requiring that the signal voltage not persist for more than 800ns after initially exceeding threshold.  It serves to filter out long-lived anthropogenic noise sources.  
Because of this criterion, only 49\% of triggering monopoles are rejected in the offline analysis at $\gamma=10^{7}$, but fully 94\% are rejected at $\gamma=10^{12}$.  

\subsection{ Systematic Errors}
\label{sec:systematics}

Preliminary simulations were used to select maximum impact parameters $r_m$ such that the effect of ignoring more distant monopoles would be negligible.  Also, because the value $\epsilon_s$ is used in place of the raw efficiency $\epsilon_t$, our limits are less prone to being artificially tightened by statistical fluctuations (although this conservative choice does weaken them by as much as 60\% in the case of $\gamma=10^{12}$).  

\begin{table*}
\begin{tabular}{|l|c|c|}
\hline
Simulation & \#detec. & $\frac{\sigma}{\text{km}^2}$ \\
\hline
Original & 24 & 0.66 \\
Voltage amplification reduction & 21 & 0.57 \\
Attenuation length reduction & 26 & 0.73 \\
Signal phase shifting & 21 & 0.57 \\
Energy loss increase & 19 & 0.50 \\
Total initial energy decrease & 20 & 0.54 \\
\hline
\end{tabular}
\caption{Error analysis simulations, the number of detections in each (out of 10000 at $\gamma=10^{10}$ using $r_m=20$ km), and the consequent cross section for monopole detection.}
\label{tableerror}
\end{table*}

Inaccurate modelling of voltage amplification, radio signal attenuation in ice, signal phase, monopole energy loss, and initial monopole energy all provide potential contributions to systematic error.  In order to quantify these errors, the simulation was modified and rerun 5 times at the intermediate $\gamma=10^{10}$.  A summary of our systematic checks is presented in Table \ref{tableerror}.

First, we reanalyzed our sample of 10k monopole energy loss profiles at $\gamma=10^{10}$, modifying each of several RadioMC parameters one by one.

Reducing the RICE amplifiers' voltage amplification by a factor of 2 in each channel does not cause a statistically-significant (at the 95\% level) change in detection sensitivity.  
This is expected because most detected monopoles produce signals well in excess of the triggering thresholds.

Reducing the in-ice signal attenuation length by a factor of 2 at all frequencies also causes a statistically insignificant change in sensitivity.  
Increasing signal attenuation causes some otherwise detectable monopoles to become too faint to trigger, but it also allows some monopoles that would otherwise fail the time-over-threshold cut to pass.  These competing effects largely cancel each other out.  

To explore the effects of variation in signal phase, we consider the results of subjecting each subshower's contribution to a phase shift of $\left(\frac{\pi}{2}\right)\left(\frac{\Delta\theta}{\sigma_\theta}\right)$, where $\Delta\theta$ is the difference between the viewing angle and Cherenkov angle and $\sigma_\theta$ is the frequency-dependent half-width (divided by a factor of 1.17) of the Cherenkov radiation's angular distribution\cite{Alvarez00}.  This is a large, ad-hoc phase variation that we used to test the potential effects of phase shifts on sensitivity.  Again, no statistically significant change in detector sensitivity occurs.  This indicates that our results are relatively insensitive to phase variations with viewing angle.

Next, we generated a new sample of 10k monopole energy loss profiles, assuming a factor of 10 increase in monopole energy loss.  This extreme change causes a modest (and statistically insignificant at the 95\% level) decrease in monopole detections.  From Fig. \ref{monobyparts}, it can be seen that the results of raising energy loss by a factor of 10 will be similar to the results of increasing $\gamma$ by a factor of about 8, so the small magnitude of this change is expected in light of the earlier results (Fig. \ref{figflux}).

Finally, we generated a second new sample of 10k monopole energy loss profiles, assuming a factor of 100 decrease in initial monopole energy.  This lower initial energy is more consistent with, for example, the initial energy assumed by the Baikal Collaboration for their analysis\cite{Aynutdinov05}.  
The resulting sensitivity decrease is statistically insignificant.  Lowering total energy at fixed $\gamma$ is equivalent to decreasing the monopole's rest mass, and energy loss is only weakly mass dependent (Fig. \ref{energyloss}).  While decreasing initial energy may sometimes cause more monopoles to be trapped in the Earth, this effect makes no difference at $\gamma=10^{10}$, for which the Earth is opaque to monopoles in either case.

In the above discussion, we have assumed a much greater uncertainty in energy loss and initial energy than in signal attenuation or amplification.  In general, RICE's sensitivity to relativistic monopoles is largely determined by the detector/ice geometry, with details such as energy loss and signal phase shift models playing much smaller roles.  The photonuclear component of monopole energy loss is the least well-established aspect of the energy loss model.  
Although variations by an order of magnitude are seen in the literature (Sec. \ref{sec:model}), these variations are unlikely in any case to alter our limits by more than a factor of 2.

\section{Summary}

From the nonobservation of highly ionizing shower ``trails'' we have derived the monopole flux upper limits shown in Fig. \ref{figflux}, which are on the order of $10^{-18} (\text{cm}^2\text{ s sr})^{-1}$.  Previously, AMANDA\cite{Wissing07}, Baikal\cite{Aynutdinov05}, and MACRO\cite{Ambrosio02} determined monopole flux limits on the order of $10^{-16}(\text{cm}^2\text{ s sr})^{-1}$ for $\beta$ greater than $0.8$, $0.8$, and $4\times10^{-5}$, respectively.  Although the results of this study cover a much narrower range of $\beta$ values than previous works, it is the range that is of the greatest interest for IMM searches.  Within much of this kinematic range ($E=10^{16}$ GeV; $\gamma\geq10^{8}$), monopole flux limits from RICE are stronger than the limits from any previous astrophysical monopole search by more than an order of magnitude.

\medskip 

 {\bf Acknowledgments}
We acknowledge helpful conversations with Alfred Goldhaber, Tom Weiler, and Stuart Wick.  This research was supported by the University of Kansas and by the National Science Foundation under grants No. OPP-0338219 and No. PHY-0243935.  JPR is supported by DOE-HEP Grant No. DE-FG02-04ER14308.


\begin{thebibliography}{99}
\bibitem{Baikalnote}The flux limit from Baikal assumes an initial monopole energy of $10^{12}-10^{15}$ GeV, lower than the initial energy assumed in this study.  Therefore, the Baikal limit shown in Fig. \ref{figflux} should be regarded as a function of the indicated mass but not necessarily the indicated $\gamma$.
\bibitem{notebeta}The loss symbol $\beta$ should cause no confusion with the velocity $\beta$.
\bibitem{noteerror}We find that Fig. 2 of \cite{Wick03} underreports the photonuclear energy loss by a $\gamma$-dependent factor of 3 to 10, due to inappropriate binning for numerical integration.  In fact, pair production is never the dominant energy loss mechanism for monopoles of rest mass 100 TeV.
\bibitem{noteorigin}More precisely, a monopole's impact parameter is calculated relative to an origin on the surface above the RICE antennas.
\bibitem{noterange}In practice, our results are insensitive to the path cutoff values, and would be essentially identical if symmetric segments, such as 450 m in front and 450 m behind, were used.
\bibitem{notebound}No simulated monopoles at $\gamma=10^6$ caused triggers.  The restrictions on $\gamma=10^6$ flux limits that could be placed with RICE, as reported in Table \ref{tableflux}, are found by assuming 2.995 detected Monte Carlo monopoles in the described simulation series and using the raw efficiency in place of the Agresti-Coull lower bound.
\end{thebibliography}
\end{document}